\documentclass[twocolumn,prl,longbibliography,superscriptaddress,floatfix]{revtex4-1}

\usepackage{graphicx}
\usepackage{amsmath}
\usepackage{amssymb}
\usepackage{bm}
\usepackage{times}

\usepackage[pdftex]{hyperref}
\hypersetup{colorlinks=true,linkcolor=blue,citecolor=blue,urlcolor=blue}

\newcommand{\pbc}{\pi}
\newcommand{\abc}{\overline{\pi}}

\begin{document}

\title{Fractal Dimension of Interfaces in Edwards-Anderson and
Long-range Ising Spin Glasses:\\Determining the Applicability of
Different Theoretical Descriptions}

\author{Wenlong Wang}
\email{wenlongcmp@gmail.com}
\affiliation{Department of Physics and Astronomy, Texas A\&M University,
College Station, Texas 77843-4242, USA}
\date{\today}

\author{M.~A.~Moore}
\affiliation{School of Physics and Astronomy, University of Manchester, 
Manchester M13 9PL, United Kingdom}

\author{Helmut G.~Katzgraber}
\affiliation{Department of Physics and Astronomy, Texas A\&M University, 
College Station, Texas 77843-4242, USA} 
\affiliation{1QB Information Technologies (1QBit), Vancouver, British
Columbia, Canada V6B 4W4}
\affiliation{Santa Fe Institute, 1399 Hyde Park Road, Santa Fe,
New Mexico 87501, USA}

\begin{abstract}

The fractal dimension of excitations in glassy systems gives information
on the critical dimension at which the droplet picture of spin glasses
changes to a description based on replica symmetry breaking where the
interfaces are space filling. Here, the fractal dimension of domain-wall
interfaces is studied using the strong-disorder renormalization group
method pioneered by Monthus [Fractals {\bf 23}, 1550042 (2015)] both for
the Edwards-Anderson spin-glass model in up to $8$ space dimensions,
as well as for the one-dimensional long-ranged Ising spin-glass with
power-law interactions. Analyzing the fractal dimension of domain walls,
we find that replica symmetry is broken in high-enough space dimensions.
Because our results for high-dimensional hypercubic lattices are limited
by their small size, we have also studied the behavior of the
one-dimensional long-range Ising spin-glass with power-law interactions.
For the regime where the power of the decay of the spin-spin
interactions with their separation distance corresponds to 6 and higher
effective space dimensions, we find again the broken replica symmetry
result of space filling excitations. This is not the case for smaller
effective space dimensions. These results show that the dimensionality
of the spin glass determines which theoretical description is
appropriate. Our results will also be of relevance to the Gardner
transition of structural glasses.

\end{abstract}

\pacs{75.50.Lk, 75.40.Cx, 05.50.+q}

\maketitle

Spin glasses have been studied for more than half a century but there is
still no consensus as to what order parameter describes their
low-temperature phase. There are two competing theories: The oldest is
the replica symmetry breaking (RSB) theory of Parisi
\cite{parisi:79,parisi:83,rammal:86,mezard:87,parisi:08}, which is known
to be correct for the Sherrington-Kirkpatrick (SK) model
\cite{sherrington:75}, which is the mean-field or infinite-dimensional
limit of the short-range Edwards-Anderson (EA) Ising spin-glass model
\cite{edwards:75}, the commonly used model for $d$-dimensional systems.
Within the RSB picture there are a very large number of pure states. In
a second theory, known as the ``droplet'' picture
\cite{mcmillan:84,bray:86,fisher:88} there are only two pure states and
the low-temperature state is replica symmetric. In the droplet picture
the behavior of the low-temperature phase is determined by low-lying
excitations or droplets whose (free) energies scale in their linear
extent $\ell$ as $\ell^{\theta}$ and whose interfaces have a fractal
dimension $d_s < d$. In the RSB theory, however, there exist low-lying
excitations which cost an energy of ${O}(1)$ and which are
space filling, that is, $d_s=d$. It has been argued \cite{moore:11} that
when $d \le 6$ the droplet picture applies while for $d > 6$ RSB is the
appropriate picture. Note, however, that in finite space dimensions RSB
is different from its infinite-dimensional limit; see Newman and Stein
\cite{newman:98,newman:01b,newman:02}, as well as Read \cite{read:14a}
for details. In this paper we study the fractal dimension as a function
of the space dimension, $d_s(d)$ \cite{comment:theta}, to find the space
dimension at which the droplets become space-filling, i.e., when
$d_s(d)=d$. Our results are consistent with $6$ being the critical
dimension. It is, of course, difficult to overcome finite-size effects
in numerical work near $6$ dimensions. Therefore, our main evidence that
$6$ is the critical dimension comes from our study of the
one-dimensional long-range spin-glass model introduced by Kotliar,
Anderson and Stein (KAS) \cite{kotliar:83}. The calculational technique
which we have used is the strong-disorder renormalization group (SDRG)
introduced by Monthus \cite{monthus:15}. This approach produces
estimates of $d_s$, that are in agreement with results on the EA model
using other numerical techniques for space dimensions $2$ and $3$ (also
studied by Monthus in Ref.~\cite{monthus:15}). In this Letter, we extend
the results of Ref.~\cite{monthus:15} up to $d = 8$ space dimensions,
and apply the method introduced in the aforementioned reference to the
KAS spin-glass model \cite{kotliar:83}.

Whether there is RSB or not in dimensions $d \le 6$ is not only
important for spin glasses. In structural glasses there has been much
recent interest in the Gardner transition, which is the transition at
which replica symmetry breaking is supposed to occur to a glass state of
marginal stability (for a review see Ref.~\cite{charbonneau:14}).
However, recent numerical results have suggested that fluctuation
effects about the mean-field solution might destroy the Gardner
transition in at least $3$ space dimensions \cite{scalliet:17}. This
result is entirely consistent with our expectation that replica symmetry
breaking will be absent for $d \le 6$.

The Edwards-Anderson model \cite{edwards:75} is defined on a
$d$-dimensional cubic lattice of linear extent $L$ by the Hamiltonian
\begin{equation} 
\mathcal{H} = - \sum_{\langle ij \rangle} J_{ij} S_i S_j, 
\label{eq:ham} 
\end{equation} 
where the summation is over only nearest-neighbor bonds and the random
couplings $J_{ij}$ are chosen from the standard Gaussian distribution
of unit variance and zero mean. The Ising spins take the values $S_i
\in \{\pm 1\}$ with $i = 1,2, \ldots, L^d$.

We have studied this model in space dimensions $d=4, \ldots, 8$ using
the SDRG method \cite{monthus:15}. Reference \cite{monthus:15} studied
the cases of $d=2$ and $3$. The SDRG approach successively traces out
the spin whose orientation is most dominated by a single large
renormalized bond to another spin; when the spin is eliminated the
couplings of the remaining spins are renormalized accordingly. We refer
the reader to Ref.~\cite{monthus:15} for further details.

The observable we focus on is related to the bond average of
$\Sigma^{\rm DW}$, where $\Sigma^{\rm DW}$ is the number of bonds
crossed by the domain wall when the boundary conditions
in one direction are changed from periodic to antiperiodic. The SDRG
method is essentially a way of constructing a possible ground state of
the system. One runs the method twice, first with periodic and next
with antiperiodic boundary conditions in one direction, and counts the
bonds across which the relative spin orientation across the bond has
altered because of the change of boundary conditions. Pictures of a
domain wall so constructed for dimension $d=2$ can be found in
Ref.~\cite{monthus:15}. It wanders, indicating that it has a
fractal dimension and its length can be described by a fractal exponent
$d_s$, where $\Sigma^{\rm DW} \sim L^{d_s}$. If the interface were
straight across the system, its length would be proportional to
$L^{d-1}$. This means that because of the wandering one expects that
$d_s > d-1$. In the RSB phase the domain walls are space filling, i.e.,
$d_s=d$. In general, $d-1 \le d_s \le d$.

We first introduce a more formal definition of $\Sigma^{DW}$ which has a
natural extension when we study long-range systems when the definition
of an interface is far from obvious. One defines the link overlap
\cite{hartmann:02} via
\begin{equation}
q_{\ell} =\frac{1}{N_b} 
\sum_{\langle ij \rangle} 
S_i^{(\pbc)}S_j^{(\pbc)} 
S_i^{(\abc)}S_j^{(\abc)} 
(2 \delta_{J_{ij}^{\pbc},J_{ij}^{\abc}} - 1).
\label{eqn:pqdef}
\end{equation}
Here $S_i^{(\pbc)}$ and $S_i^{(\abc)}$ denote the ground states found
with periodic (${\pbc}$) and antiperiodic (${\abc}$) boundary
conditions, respectively. One can switch from periodic to antiperiodic
boundary conditions by flipping the sign of the bonds crossing a
hyperplane of the lattice. $N_b$ is the number of nearest-neighbor
bonds in the lattice which for a $d$-dimensional hypercube is given by
$N_b=d L^d$. One can then define \cite{hartmann:02}
\begin{equation}
\Gamma \equiv 1-q_{\ell}= \frac{2\Sigma^{\rm DW}}{d L^d} \sim L^{d_s-d}\, .
\label{eqn:gammadef}
\end{equation}

\begin{figure}[htb]
\begin{center}
\includegraphics[width=\columnwidth]{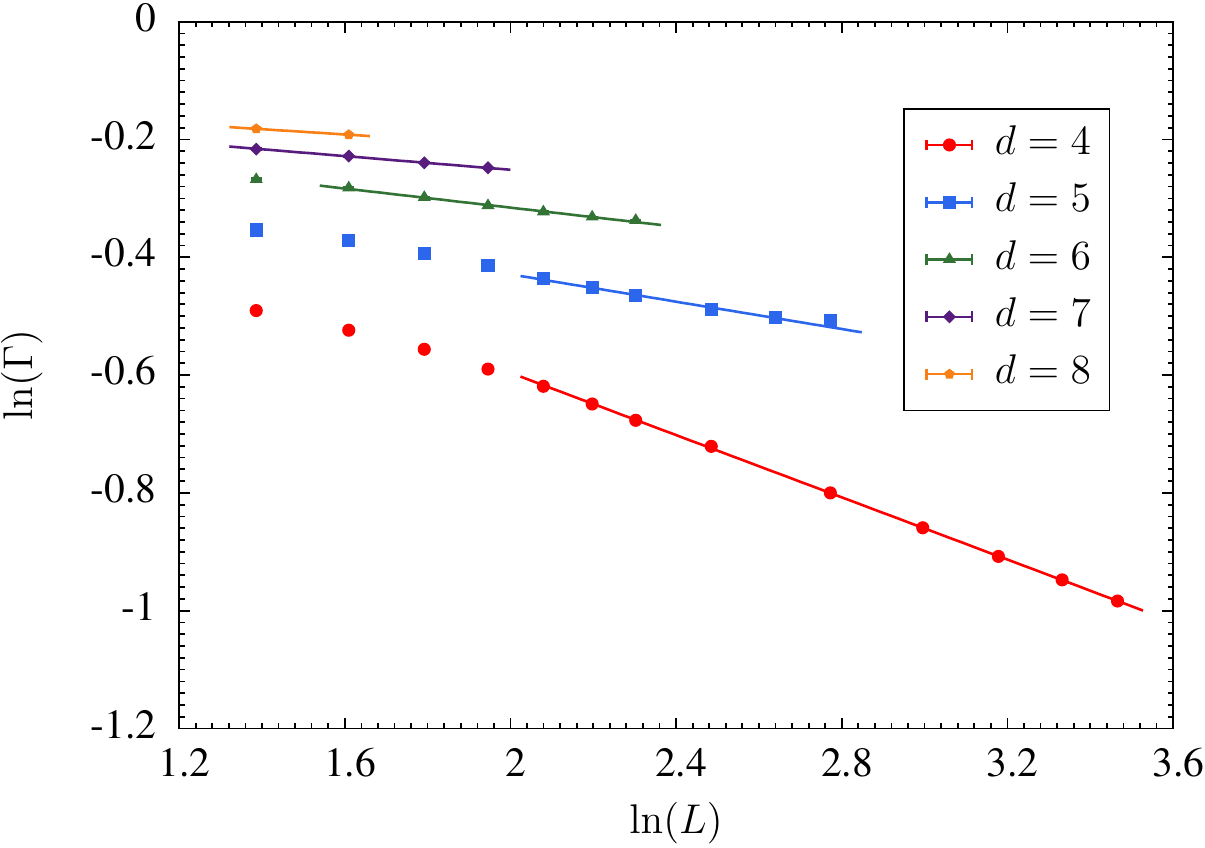}
\caption{
$\Gamma$ [see Eq.~\eqref{eqn:gammadef}] for various dimensions $d$ for the EA
model as a function of their linear dimension $L$. Note that $\Gamma
\sim L^{d_s-d}$. Our estimate of $d_s$ is determined by the slope of the
straight lines drawn through the points at large $L$ values. Error bars
are smaller than the symbols.
}
\label{EASIGMA}
\end{center}
\end{figure}

\begin{figure}[htb]
\begin{center}
\includegraphics[width=\columnwidth]{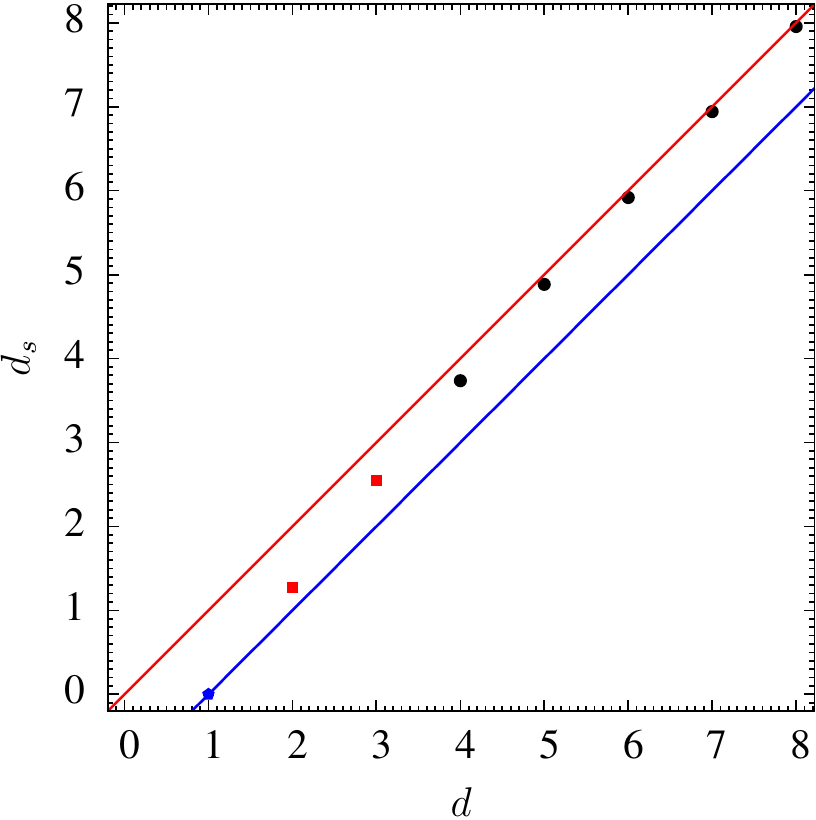}
\caption{
Values of the fractal dimension $d_s$ as a function of the space
dimension $d$ determined using the SDRG method. The top (red) line is
the upper bound where $d_s=d$ and the bottom (blue) line is the lower
bound where $d_s=d-1$. The value for $d=1$ (blue pentagon) can be
calculated analytically. The values for $d = 2$ and $3$ (red squares)
are taken from Ref.~\cite{monthus:15}. The statistical error bars are 
smaller than the symbols. No systematic errors have been considered.
}
\label{EAds}
\end{center}
\end{figure}

In Fig.~\ref{EASIGMA} we show the bond-averaged value of $\Gamma$
[Eq.~\eqref{eqn:gammadef}] vs $\ln L$ which should be a straight line of
slope $d_s-d$. In Fig.~\ref{EAds} the value of $d_s$ is plotted for
various dimensionalities $d$. For $d=1$, $d_s(1) = 0$ (pentagon), while
for $d=2$ we have used the value from Ref.~\cite{monthus:15}, i.e.,
$d_s(2) = 1.27$ (square), which is in excellent agreement with other
numerical estimates
\cite{bray:87,middleton:01,hartmann:02,melchert:07,amoruso:06a,bernard:07,risau-gusman:08}.
For $d=3$, Ref.~\cite{monthus:15} quotes $d_s(3) = 2.55$ (square), which
is again in good agreement with other estimates
\cite{palassini:00,katzgraber:01}. In addition, we estimate $d_s(4) =
3.7358(13)$, which again is in good agreement with Monte Carlo estimates
\cite{katzgraber:01}. Note that the largest system in
Ref.~\cite{katzgraber:01} has $N = 5^4$ spins, which seems to not be in
the scaling regime (see Fig.~\ref{EASIGMA}). This means that results
from small systems tend to overestimate $d_s$.

Finally, one can see that as the dimensionality $d$ increases, $d_s(d)$
approaches $d$. However, results from simulations on hypercubic lattices
struggle from corrections to scaling. These make it difficult to claim
that $d_s = d$ at precisely $d = 6$. To address this point, we turn to
the KAS model.

\begin{table}
\caption{
Size and number of disorder realizations used in the SDRG approach for
the EA and KAS models. $d$ is the space dimension, $L$ is the linear
system size and $M$ is the number of disorder realizations used for the
average. For the KAS model, we use $M = 3000$ disorder realizations for
each $L=256$, $512$, $1024$, $2048$, $4096$, and $8192$ at the following
$\sigma$ values: $0.1$, $0.25$, $0.5$, $0.55$, $0.6$, $0.667$, $0.75$,
$0.896$, $1$, $1.25$, $1.5$, $1.75$, $2$, $2.25$, $2.5$, $2.75$, and
$3$.
\label{table}
}
\begin{tabular*}{\columnwidth}{@{\extracolsep{\fill}} c c c}
\hline
\hline
$d$ &$L$ &$M$ \\
\hline
$4$ &$\{4, 5, 6, 7, 8, 9, 10, 12, 16, 20, 24\}$ &$3000$ \\
$4$ &$28$ &$717$ \\
$4$ &$32$ &$121$ \\

$5$ &$\{4, 5, 6, 7, 8, 9, 10, 12\}$ &$3000$ \\
$5$ &$14$ &$1342$ \\
$5$ &$16$ &$581$ \\

$6$ &$\{4, 5, 6, 7, 8\}$ &$3000$ \\
$6$ &$9$ &$1843$ \\
$6$ &$10$ &$938$ \\

$7$ &$\{4, 5, 6\}$ &$3000$ \\
$7$ &$7$ &$512$ \\

$8$ &$\{4, 5\}$ &$3000$ \\
\hline
\hline
\end{tabular*}
\end{table}

The one-dimensional KAS model \cite{kotliar:83} is described by the
Hamiltonian in Eq.~(\ref{eq:ham}), except that the $L$ spins lie on a
ring and the exchange interactions $J_{ij}$ are 
long ranged, i.e., $\langle ij \rangle$
denotes a sum over {\em all} pairs of spins:
\begin{equation}
J_{ij}=c(\sigma,L) \frac{\epsilon_{ij}}{r_{ij}^{\sigma}},
\label{eqn:Jij}
\end{equation}
where $r_{ij}$ is the shortest circular length between sites $i$ and $j$ 
\cite{monthus:14b}.

The disorder $\epsilon_{ij}$ is chosen from a Gaussian
distribution of zero mean and standard deviation unity, while the
constant $c(\sigma,L)$ in Eq.~(\ref{eqn:Jij}) is fixed to make the
mean-field transition temperature $T_c^{\mathrm{MF}}=1$ and
$(T_c^{\mathrm{MF}})^2= \sum_j [J_{ij}^2]_{\mathrm{av}}$, where
$[\cdots]_{\rm av}$ represents a disorder average and $[J_{ij}^2]_{\rm
av}=c^2(\sigma,L)/r_{ij}^{2 \sigma}$ where $1/c(\sigma,L)^2=
\sum_{j=2}^L 1/r_{1j}^{2 \sigma}$. Note that in the limit $\sigma
\rightarrow 0$ the KAS model reduces to the infinite-range SK model.
The advantage of the KAS model is that one can study a large range of
linear system sizes. 

The KAS model can be taken as an interpolation between the $d = 1$ EA
model and the $d = \infty$ SK model as the exponent $\sigma$ is varied.
The phase diagram of this model in the $d$--$\sigma$ plane has been
deduced from renormalization group arguments in
Refs.~\cite{bray:86b,fisher:88,katzgraber:03}. For $0 \le \sigma <1/2$
it behaves just like the infinite-range SK model. When $1/2< \sigma
<2/3$ the critical exponents at the spin-glass transition are mean-field
like, and this corresponds in the EA model with space dimensions above
$6$. In the interval $2/3 \le \sigma <1$ the critical exponents are
changed by fluctuations away from their mean-field values. When $\sigma
\ge 1$, $T_c(\sigma)=0$ and when $\sigma > 2$, the long-range
zero-temperature fixed point, which controls the value of the exponents
$d_s$ and $\theta$, becomes identical to that of the nearest-neighbor
one-dimensional EA model, i.e., $d_s=0$ and $\theta=-1$. There is a
convenient mapping between $\sigma$ and an effective dimensionality
$d_{\textrm{eff}}$ of the short-range EA model
\cite{katzgraber:03,katzgraber:09b,leuzzi:09,banos:12b,aspelmeier:16}.
For $1/2 < \sigma < 2/3$, it is
\begin{equation}
d_{\textrm{eff}}=2/(2\sigma-1).
\label{eqn:def} 
\end{equation}
Thus, right at the value of $\sigma =2/3$, $d_{\textrm{eff}}=6$. The
arguments given in Ref.~\cite{moore:11} that the critical dimension is
$6$, below which one sees droplet behavior and above which one sees RSB
behavior were directly extended to the KAS model and predicted that only
in the interval $\sigma < 2/3$ will one see RSB behavior, so that
$\sigma =2/3$ is the critical value expected for the KAS model.

We have determined $d_s$ for the KAS model from two definitions
of $d_s$. The first definition is via the generalization of the link
overlap in Eq.~(\ref{eqn:pqdef}) to the long-range KAS model just as
done in Ref.~\cite{katzgraber:03f}:
\begin{equation}
q_{\ell} = 
\frac{2}{L(L-1)} 
\sum_{i<j} w_{ij} 
S_i^{(\pbc)}S_j^{(\pbc)} 
S_i^{(\abc)}S_j^{(\abc)} 
(2 \delta_{J_{ij}^{\pbc},J_{ij}^{\abc}} -1),
\label{eqn:pqdeflr}
\end{equation}
where $w_{ij}=(L-1) c(\sigma,L)^2/r_{ij}^{2\sigma}$. Note that the sum
of $w_{ij}$ over $i<j$ equals $L(L-1)/2$. Antiperiodic boundary
conditions can be produced by flipping the sign of the bonds when the
shortest paths go through the origin. $d_s$ is then obtained from
$q_{\ell}$ using Eq.~(\ref{eqn:gammadef}) with $d=1$.

Because we are unsure of the topological significance of $d_s$
calculated in this way, we use a second approach whose topological
significance is clear. Fortunately, it gives very similar results to that
of our first definition. Let $\tau_i=S_i^{(\pi)}
S_i^{(\overline{\pi})}$, and define an ``island'' as a sequence in which
all the $\tau_i$ are of the same sign. For the EA model limit of the
KAS model, i.e., when $\sigma > 2$, there are only two islands but when
the long-range zero-temperature fixed point \cite{bray:86b} controls the
behavior, there are many islands; we denote by $N_I$ the number of
islands produced by the change from periodic to antiperiodic boundary
conditions. Formally, $N_I$ can be computed via
\begin{equation}
N_I= \frac{1}{4} \sum\limits_{i=1}^{L} (\tau_{i+1}-\tau_i)^2,
\end{equation}
where $\tau_{L+1}=\tau_1$. We define $d_s$ via
$N_I \sim L^{d_s}$.
The islands have a distribution of sizes with their mean size $L_0 =
L/N_I \sim L^{1-d_s}$. In the RSB region where $d_s =d =1$ $L_0$ is
independent of the size of the system and is of ${O}(1)$, a
result which we obtained previously from direct studies in the SK limit
\cite{aspelmeier:16a}.

\begin{figure}[htb]
\begin{center}
\includegraphics[width=\columnwidth]{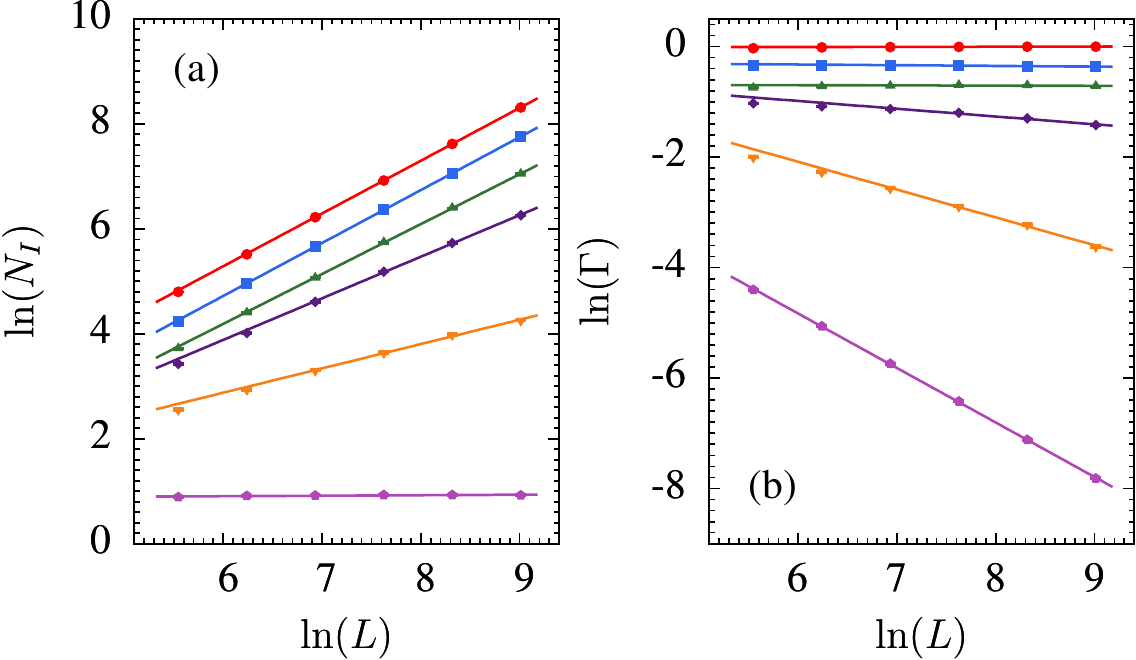}
\caption{
Dependence of $N_I $ (expected form $\sim L^{d_s})$ and $\Gamma$
(expected form $\sim L^{d_s-1})$ on $L$ for the KAS model obtained via
the SDRG method for a few representative values of the exponent
$\sigma$: $0.1$, $0.5$, $0.667$, $0.75$, $1.0$, and $2.0$. In both
(a) and (b) the values of $\sigma$ increase from top to bottom.
Error bars are smaller than the symbols.}
\label{KAS}
\end{center}
\end{figure}

\begin{figure}[htb]
\begin{center}
\includegraphics[width=\columnwidth]{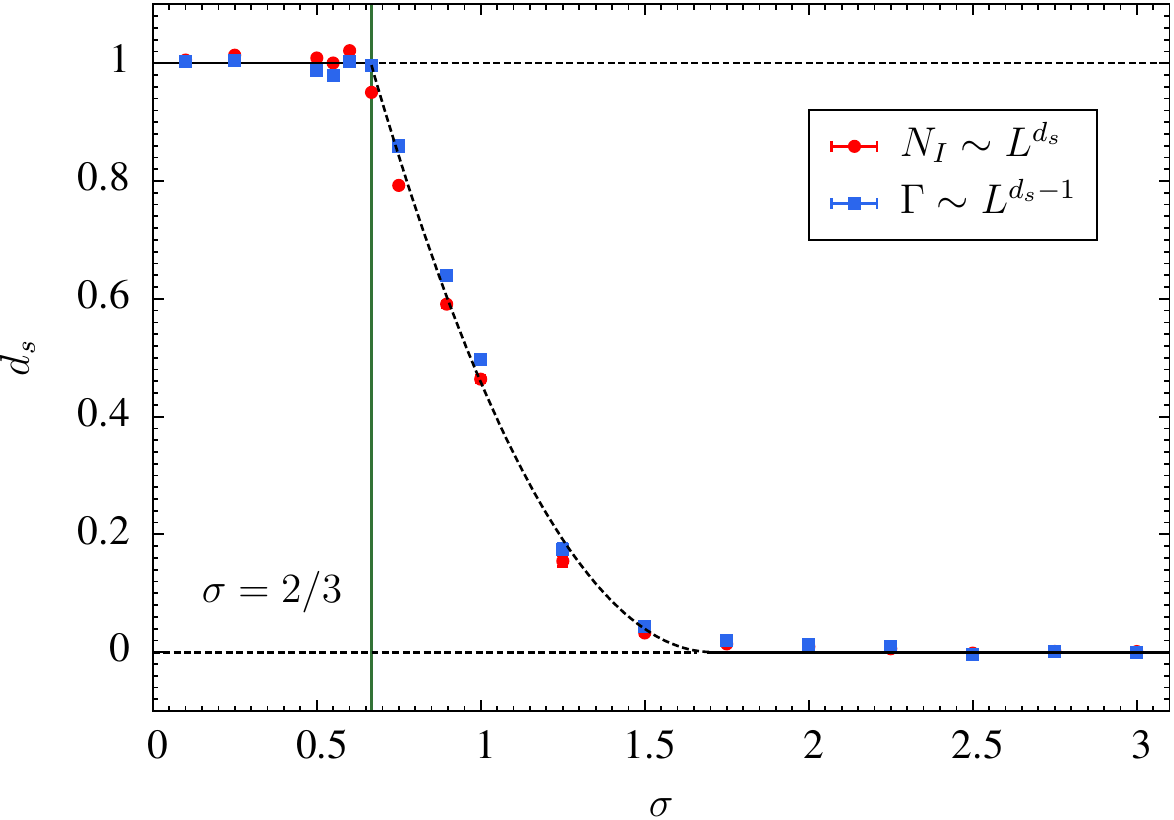}
\caption{
Exponent $d_s$ calculated from the scaling of $N_I$ (red circles) and
$\Gamma$ (blue squares) using the SDRG method at all the $\sigma$ values
which we have studied (see Table \ref{table}). The vertical (green) line
marks where $\sigma =2/3$, which is the value of $\sigma$ at
which we expect $d_s$ to decrease below unity. The black or dashed lines are
guides to the eye; the dashed line interpolating $d_s=1$ and $d_s=0$ is a cubic fit.
The values for $d_s$ are obtained by fitting a straight line to the four
largest system sizes studied ($L = 1024$, $2048$, $4096$, and $8192$),
except for values of $\sigma\ge 1.5$, where all system sizes are used
for the fits. 
Error bars are smaller than the symbols.}
\label{KAS5}
\end{center}
\end{figure}

We have used these two quite distinct definitions of $d_s$ to compute
the fractal dimension as a function of $\sigma$ using the SDRG method.
The details of the system sizes and numbers of disorder realizations
can be found in Table \ref{table}. Our results for $N_I$ and $\Gamma$
 are shown in Fig.~\ref{KAS}. From these we have extracted values for
$d_s$ which are shown in Fig.~\ref{KAS5}. The values obtained for $d_s$
from $\Gamma$ and $N_I$ are reassuringly similar. The most striking
feature of our results are, first, $d_s \simeq 1(=d) $ when $\sigma <
2/3$, and second, $d_s$ decreases from unity as $\sigma$ increases past
$2/3$. Because $\sigma =2/3$ maps to $d=6$ according to
Eq.~(\ref{eqn:def}) we believe that this is strong evidence that $6$ is
the dimension below which the droplet picture applies and that only in
more than $6$ space dimensions will one find RSB effects, just as
anticipated in Ref.~\cite{moore:11}.

At $\sigma > 2$ the long-range fixed point is unstable and the
renormalization group flows go to the short-range fixed point, that of
the $d=1$ EA model \cite{bray:86b}. For the EA model in one space
dimension, $d_s=0$ and $\theta =-1$. We were expecting that $d_s$ would
go to zero at $\sigma =2$; it is possible that $d_s$ is just very small
in the interval $1.5 < \sigma < 2$.

There are small finite-size corrections when using the SDRG method. For
$\sigma \sim 1$, there is a downward curvature in the data
(Fig.~\ref{KAS}) so that if we had been able to study larger $L$ values,
our estimates of $d_s$ might have decreased. However, the behavior in
the crucial region where $\sigma$ is close to $2/3$ is less affected by
finite-size effects. Monthus and Garel \cite{monthus:14c} have obtained
estimates for $d_s$ from exact studies on the KAS model for $L \le 24$.
They found $d_s(\sigma =0.62) \simeq 1$, $d_s(\sigma =0.75)\simeq 0.94$,
$d_s(0.87) \simeq 0.82$, $d_s(\sigma=1) \simeq 0.72$, and
$d_s(\sigma=1.25) \simeq 0.4$. These results illustrate clearly that
estimates of $d_s$ from small systems tend to be high.

We now discuss the accuracy of the SDRG method. First, we note that SDRG
is considerably better than the Migdal-Kadanoff (MK) approximation which
gives $d_s^{{\rm MK}} = d-1$ \cite{monthus:15}, which coincides with the
lower bound on $d_s$ and so never gives $d_s=d$. The SDRG method can be
used to determine $\theta$ as well as $d_s$. In $2$ dimensions; it gives
$\theta \simeq 0$ \cite{monthus:15}; its established value is close to
$-0.28$ \cite{hartmann:02c}. The SDRG method is only exact for special
cases. Like the MK approximation, it is exact in one space dimension for
the EA model but its performance for the energy per spin and the
exponent $\theta$ then steadily deteriorates with increasing space
dimension $d$.  Monthus \cite{monthus:15} suggested that it does a good
job for the exponent $d_s$ because that exponent is dominated by short
length-scale optimization which is well captured by the early steps of
the SDRG method, but that it does badly for the interface free-energy
exponent $\theta$ which also requires optimization on the longest length
scales. We also suspect that its success in determining $d_s$ might be
connected with the fact that the domain wall is a self-similar fractal.
That means it has the same fractal dimension $d_s$ whether that fractal
dimension is studied on short or long length scales. In $d=2$ and $d=3$
Monthus \cite{monthus:15} showed that the SDRG worked on short length
scales but fails on long length scales. We believe the consequence of
this might just be that in determining the length of the domain wall
$\Sigma^{\rm DW} =A L^{d_s}$, the exponent $d_s$ is correctly determined
from the short length-scale behavior, but to obtain the coefficient $A$
correctly one would need a treatment also valid on long length scales.
In the KAS model at $\sigma =0.1$ the SDRG fails on short length scales
but works on long length scales. Again, we believe that the exponent
$d_s=d=1$ is correct, but that the coefficient $A$ is only approximate.

One worrisome issue is that numerical work around $6$ space dimensions
could suffer from poor precision, so how can one be confident that $d=6$
is a special space dimension below which RSB does not occur (aside from
a rigorous proof). There is another numerical procedure, the greedy
algorithm \cite{cieplak:94,newman:94,jackson:10,sweeney:13} in which one
satisfies the bonds in the order of the couplings $\vert J_{ij}\vert$
unless a closed loop appears, where one skips to the next largest bond.
We have found that as $d \to 6$ from below the values of $d_s$ obtained
from the GA approach those from the SDRG, which is not surprising when
one examines how the SDRG works. For $d=2$, however, the GA is certainly
poorer than the SDRG, because it predicts $d_s = 1.216(1)$
\cite{sweeney:13}. Jackson and Read \cite{jackson:10}, however, have an
analytical argument that $6$ is a special space dimension for the GA
algorithm. This gives us confidence that $6$ is the space dimension
above which interfaces are space filling.

\begin{acknowledgments}

W.~W.~and H.~G.~K.~acknowledge support from NSF DMR Grant No.~1151387. The
work of H.G.K. and W.W is supported in part by the Office of the
Director of National Intelligence (ODNI), Intelligence Advanced Research
Projects Activity (IARPA), via MIT Lincoln Laboratory Air Force Contract
No.~FA8721-05-C-0002. The views and conclusions contained herein are
those of the authors and should not be interpreted as necessarily
representing the official policies or endorsements, either expressed or
implied, of ODNI, IARPA, or the U.S.~government. The U.S.~Government is
authorized to reproduce and distribute reprints for Governmental purpose
notwithstanding any copyright annotation thereon. We thank Texas A\&M
University for access to their Ada and Curie clusters.

\end{acknowledgments}

\bibliography{refs,comments}

\begin{thebibliography}{44}
\bibitem [{com()}]{comment:theta}%
  \BibitemOpen
  \href@noop {} {}\bibinfo {note} {{We note that the behavior of the stiffness
  exponent $\theta$ was discussed in detail in
  Ref.~\cite{aspelmeier:16a}.}}\BibitemShut {Stop}%
\end{thebibliography}

\end{document}